\documentclass[a4paper,11pt]{article}
\usepackage{pos}
\usepackage{soul}

\title{LST-1 observations of GRB 221009A: Insights into its late-time VHE afterglow}

\author*[a]{Arnau Aguasca-Cabot}
\author[b]{Susumu Inoue}
\author[c]{Monica Seglar-Arroyo}
\author[d]{Kenta Terauchi}
\author[a]{Pol Bordas}
\author[a]{Marc Ribó}
\onbehalf{on behalf of the CTAO-LST Project}

\affiliation[a]{Departament de Física Quàntica i Astrofísica, Institut de Ciències del Cosmos, Universitat de Barcelona, IEEC-UB, Martí i Franquès, 1, 08028, Barcelona, Spain.}

\affiliation[b]{Institute for Cosmic Ray Research, University of Tokyo, 5-1-5, Kashiwa-no-ha, Kashiwa, Chiba 277-8582, Japan.}

\affiliation[c]{Institut de Fisica d’Altes Energies (IFAE), The Barcelona Institute of Science and Technology, Campus UAB, 08193 Bellaterra (Barcelona), Spain.}

\affiliation[d]{Department of Physics, Tokai University, 4-1-1, Kita-Kaname, Hiratsuka, Kanagawa 259-1292, Japan.}

\emailAdd{arnau.aguasca@fqa.ub.edu}

\abstract{
    Gamma-ray bursts (GRBs) originate from explosions at cosmological distances, generating collimated jets. GRB 221009A, exploded on 9 October 2022, has been established as the brightest GRB to date. Its bright and long emission was extensively followed up from radio to gamma rays. LHAASO firmly detected the onset of the afterglow emission at energies up to $\sim$13 TeV within about an hour after the burst, starting just a few minutes after the trigger. While this VHE emission component can be accounted for in a narrow jet scenario, such an interpretation cannot reproduce the broadband emission observed at later times, which exceeds the theoretical expectations. This discrepancy can be settled if more complex models are considered, providing the first strong evidence for a structured jet in a long GRB. Unfortunately, the VHE emission after a few hours is poorly constrained, as sensitive VHE observations by Cherenkov Telescopes were prevented due to strong moonlight conditions. The first Large-Sized Telescope (LST-1) of the future Cherenkov Telescope Array Observatory began observations about one day after the burst under high night sky background conditions. These observations are the first ones performed on GRB 221009A by a Cherenkov telescope, revealing a hint of a signal with a statistical significance of about 4$\sigma$ during the observations performed at 1.3 days after the burst. The monitoring campaign continued until the end of November 2022, making it the deepest observation campaign performed on a GRB with the LST-1. In this contribution, we will present the analysis results of the LST-1 observation campaign on GRB 221009A in October 2022.
}

\ConferenceLogo{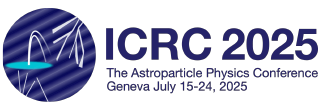}

\FullConference{39th International Cosmic Ray Conference (ICRC2025)\\
 15–24 July 2025\\
Geneva, Switzerland\\}

\begin{document}
\maketitle

\section{Introduction}

    Gamma-ray bursts (GRBs) were first detected in 1967 with the {Vela} satellites \citep{1973ApJ...182L..85K}. These events are characterised by intense {emission} (milliseconds to thousands of seconds) in the hard X-ray to soft-gamma-ray range (tens of keV to several MeV). The majority of the emitted energy during this phase occurs in the sub-MeV to MeV band and is commonly referred to as the prompt emission. A long-lived, multi-wavelength afterglow emission was first observed in 1997 \citep{1997Natur.387..783C}. This fading afterglow emission can persist for several {months} following the burst discovery and has been detected across the electromagnetic spectrum from radio to very-high-energy gamma rays (VHE; $E>100$\, GeV) \citep[e.g.][]{Ren2024}.
    
    GRBs originate from catastrophic events at cosmological distances that launch relativistic jets. A GRB is detected from Earth when one of the jets is pointing towards us. The observed afterglow emission is interpreted as non-thermal radiation produced by the interaction of relativistic particles with the surrounding medium. The physical origin of GRBs is broadly divided into two categories. Long-duration {($T>2$ s)} GRBs are associated with the core collapse of massive stars, while short-duration {($T<2$ s)} GRBs are connected to the coalescence of two compact objects \citep{2012grb..book..169H,Berger2014}.
    
    The non-thermal high-energy emission is usually attributed to synchrotron self-Compton processes. The first VHE gamma-ray detection from a GRB was reported for GRB~190114C \citep{2019GCN.23701....1M,2019Natur.575..455M}. To date, five GRBs have been detected in this band \citep{abdalla2019very,HESS190829A,Abe_2023_GRB201216C,LHAASO2023WCDA}. Among these, GRB~221009A stands out as the brightest GRB ever detected, earning the nickname of the brightest-of-all-time (BOAT) GRB. It was initially discovered on 9 October 2022 by several space-borne satellites, including \textit{Swift} and \textit{Fermi} \citep{Williams2023swift,Lesage_2023,2025ApJS..277...24A}. An extensive follow-up campaign continued across the electromagnetic spectrum over a broad period of time. In the VHE band, the Large High Altitude Air Shower Observatory (LHAASO) was serendipitously covering the sky region where GRB~221009A occurred, succeeding in detecting the onset afterglow emission \citep{LHAASO2023WCDA,LHAASO2023KM2A}.
    
    On 9 October 2022, the presence of the bright Moon prevented rapid follow-up observations by imaging atmospheric Cherenkov telescopes (IACTs). The first Large-Sized Telescope (LST-1) of the future Cherenkov Telescope Array Observatory (CTAO) started observation 1.33 days after the burst, constituting the first observation by this class of detectors. 
        
    In this work, we report the observations with LST-1. Section~\ref{sect:observations} describes the observation campaign with LST-1. The analyses of the observations are described in Sect.~\ref{sect:analysis}, while the results are shown in Sect.~\ref{sect:results}. Concluding remarks are provided in Sect.~\ref{sect:conclusions}.

\section{Observations}
\label{sect:observations}
    LST-1 began observations 1.33 days after the burst, on 10 October 2022, {with Moon illumination} at 99\%. Similar night-sky-background (NSB) conditions were present during the second day of data taking, 3.33 days post-burst, on 12 October 2022. {Observation on 11 October 2022 were not possible due to camera problems.} These observations were conducted with reduced high voltage during a period when data acquisition is normally halted due to the bright moonlight. Subsequent observations were obtained under nominal operations in low NSB\footnote{Diffuse NSB below 2.3 photoelectrons.} conditions until the end of October, within the same lunar cycle as the initial observations. The campaign continued into the following lunar cycle, extending through the end of November 2022. {Here we report} the observations under high and low NSB conditions in October 2022{, which are summarised} in Table~\ref{tab:obs_table}. 

    \begin{table}  
        \caption{Observations of GRB~221009A with \mbox{LST-1} in October 2022. For each observation day, we list the day of the evening before data taking, the starting date, the starting time offset with respect to the burst trigger \citep[$T_0$;][]{GBM2022GCN}, the observation time after the data selection, the zenith angle range of the observations, and the NSB conditions that affected the data.}
        \begin{center}
        \footnotesize
        \label{tab:obs_table}
        \begin{tabular}{lccccc}       
        \hline \hline
        \noalign{\smallskip}
        Start day & Start date &  $T-T_0$ & Time after & Zenith range & NSB conditions \\
        & & & data selection & \\
        & $\mathrm{[MJD]}$ & [d] & [h] & [\textdegree]--[\textdegree] & \\
        \noalign{\smallskip}
        \hline
        \noalign{\smallskip}
         10 Oct. 2022 & 59862.88 &  1.33 & 1.75 & 31--54 & High \\
         12 Oct. 2022 & 59864.89 &  3.33 & 1.42 & 34--52 & High\\
         15 Oct. 2022 & 59867.85 &  6.30 & 0.80 & 25--52 & Low \\
         16 Oct. 2022 & 59868.88 &  7.32 & 2.35 & 34--65 & Low \\
         17 Oct. 2022 & 59869.85 &  8.30 & 2.41 & 28--60 & Low \\
         23 Oct. 2022 & 59875.86 &  14.30\,\,\, & 2.01 & 34--61 & Low \\
         25 Oct. 2022 & 59877.89 &  16.33\,\,\, & 1.18 & 45--59 & Low \\
         26 Oct. 2022 & 59878.87 &  17.32\,\,\, & 1.42 & 42--58 & Low \\
        \noalign{\smallskip}
        \hline
        \noalign{\smallskip}
        \multicolumn{3}{c}{Total observation time after data selection [h]} & 13.34 & \\
        \noalign{\smallskip}      
        \hline        
        \end{tabular}
        \end{center}        
    \end{table}

\section{Data analysis}
\label{sect:analysis}

    We divided the observations according to the NSB conditions during the data taking. Observations on 10 and 12 October were analysed using a moon-adapted analysis scheme, while the rest of the data was analysed using the standard data analysis described in \citep{Abe_2023_LSTPerformance}.

    The moon-adapted analysis used refined camera calibrations employing an optimised charge-integration method and image cleaning adapted for an observation-by-observation analysis to account for the rapid evolving observing conditions with high NSB. These adaptations were considered after evaluating multiple combinations of charge-integration and image-cleaning algorithm. For instance, we tested different window configuration for the waveform signal integration \citep{ctapipe-icrc-2023}. Moreover, we investigated if the pulse shape of the signal between the observations in nominal and reduced high voltage significantly changed. Only minor variations, at a level of few percent were found, indicating that changes in the pulse shape are not expected to introduce significant systematic effects in the analysis. {More detailed} information about the analysis is described in \citep{GRB221009A_LST1}.

\section{Results}
\label{sect:results}

\begin{figure}[h!]
        \centering
        \includegraphics[width=0.57\linewidth]{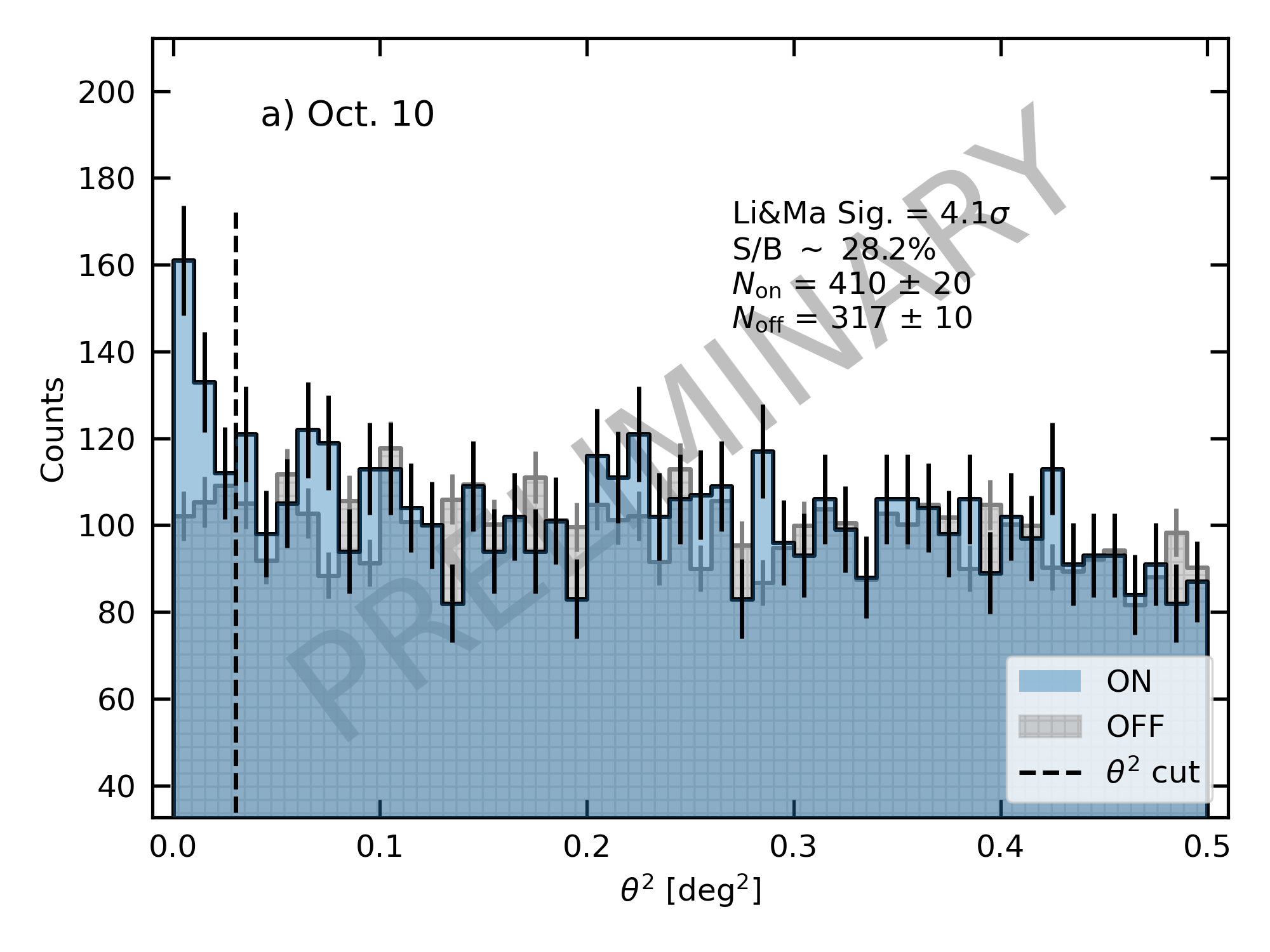}
        \includegraphics[width=0.57\linewidth]{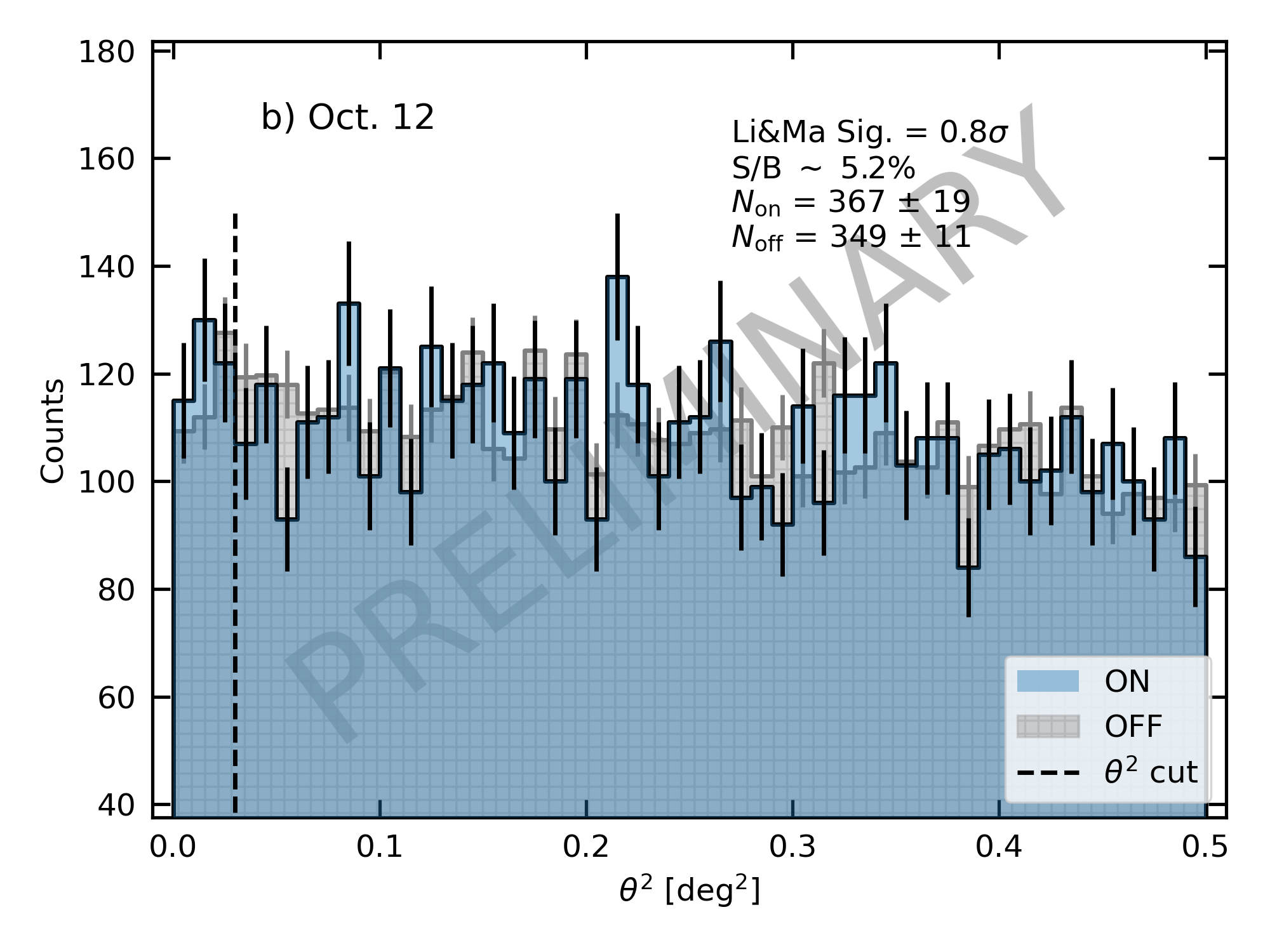}
        \includegraphics[width=0.57\linewidth]
        {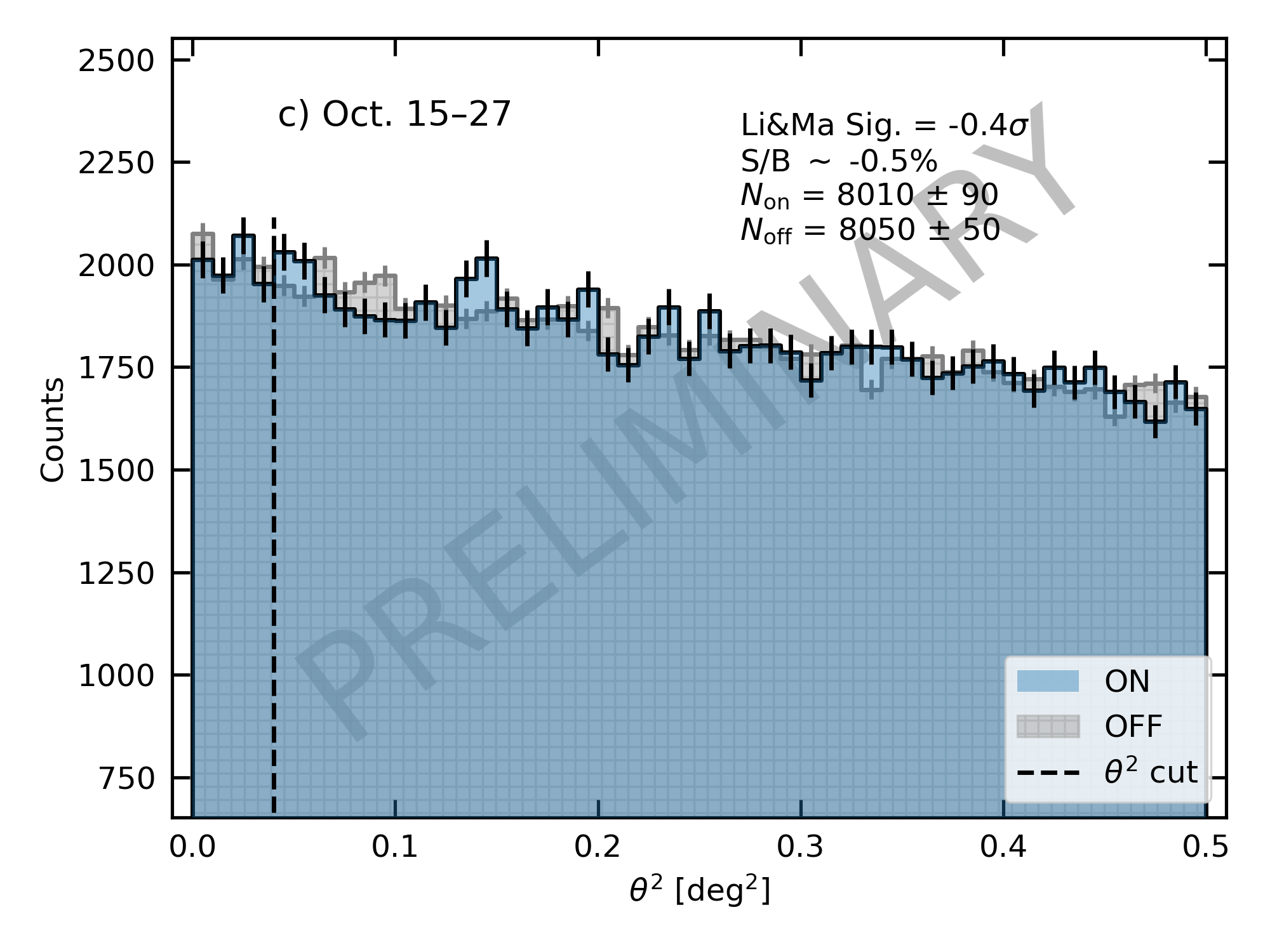}
        \caption{$\theta^2$ distribution centred on the GRB~221009A position (ON) and the mean distribution from three control (OFF) regions. Panels a), b) and c) show the distributions using all the observations on 10, 12 and 15--27 October 2022, respectively (see Table~\ref{tab:obs_table}). Error bars correspond to 1$\sigma$ statistical errors.}
        \label{fig:theta2}
    \end{figure}

    We obtain an excess of gamma rays centred towards the GRB~221009A position at a detection statistical significance of 4.1$\sigma$ on 10 October 2022, while observations on subsequent days are compatible with background. Figure~\ref{fig:theta2} shows the $\theta^2$ distribution for 10, 12 and 15--27 October 2022. The reader is referred to \citep{GRB221009A_LST1} for more information about the analysis results.

\section{Conclusions}
\label{sect:conclusions}

    LST-1 performed an extensive observation campaign on GRB 221009A. The first observations occurred 1.33 days after the burst, being the earliest observations of this burst published to date with an IACT. We obtain a hint of detection at $4\sigma$ significance on that day, while there is no excess over the expected background in the days after that one night. The dedicated analysis of these data and the final results are provided in \citep{GRB221009A_LST1}.

\bibliographystyle{JHEP}
\bibliography{bibliography}

\vspace{1cm}

{Full Author List: CTAO-LST Project}

\tiny{\noindent
K.~Abe$^{1}$,
S.~Abe$^{2}$,
A.~Abhishek$^{3}$,
F.~Acero$^{4,5}$,
A.~Aguasca-Cabot$^{6}$,
I.~Agudo$^{7}$,
C.~Alispach$^{8}$,
D.~Ambrosino$^{9}$,
F.~Ambrosino$^{10}$,
L.~A.~Antonelli$^{10}$
% ...
The CTAO-LST Project
}\\

\tiny{\noindent$^{1}${Department of Physics, Tokai University, 4-1-1, Kita-Kaname, Hiratsuka, Kanagawa 259-1292, Japan}.
$^{2}${Institute for Cosmic Ray Research, University of Tokyo, 5-1-5, Kashiwa-no-ha, Kashiwa, Chiba 277-8582, Japan}.
$^{3}${INFN and Università degli Studi di Siena, Dipartimento di Scienze Fisiche, della Terra e dell'Ambiente (DSFTA), Sezione di Fisica, Via Roma 56, 53100 Siena, Italy}.
$^{4}${Université Paris-Saclay, Université Paris Cité, CEA, CNRS, AIM, F-91191 Gif-sur-Yvette Cedex, France}.
$^{5}${FSLAC IRL 2009, CNRS/IAC, La Laguna, Tenerife, Spain}.
$^{6}${Departament de Física Quàntica i Astrofísica, Institut de Ciències del Cosmos, Universitat de Barcelona, IEEC-UB, Martí i Franquès, 1, 08028, Barcelona, Spain}.
$^{7}${Instituto de Astrofísica de Andalucía-CSIC, Glorieta de la Astronomía s/n, 18008, Granada, Spain}.
$^{8}${Department of Astronomy, University of Geneva, Chemin d'Ecogia 16, CH-1290 Versoix, Switzerland}.
$^{9}${INFN Sezione di Napoli, Via Cintia, ed. G, 80126 Napoli, Italy}.
$^{10}${INAF - Osservatorio Astronomico di Roma, Via di Frascati 33, 00040, Monteporzio Catone, Italy}.
% ...
}

\acknowledgments 
\tiny{
We gratefully acknowledge financial support from the following agencies and organisations:
Conselho Nacional de Desenvolvimento Cient\'{\i}fico e Tecnol\'{o}gico (CNPq), Funda\c{c}\~{a}o de Amparo \`{a} Pesquisa do Estado do Rio de Janeiro (FAPERJ), Funda\c{c}\~{a}o de Amparo \`{a} Pesquisa do Estado de S\~{a}o Paulo (FAPESP), Funda\c{c}\~{a}o de Apoio \`{a} Ci\^encia, Tecnologia e Inova\c{c}\~{a}o do Paran\'a - Funda\c{c}\~{a}o Arauc\'aria, Ministry of Science, Technology, Innovations and Communications (MCTIC), Brasil;
Ministry of Education and Science, National RI Roadmap Project DO1-153/28.08.2018, Bulgaria;
Croatian Science Foundation (HrZZ) Project IP-2022-10-4595, Rudjer Boskovic Institute, University of Osijek, University of Rijeka, University of Split, Faculty of Electrical Engineering, Mechanical Engineering and Naval Architecture, University of Zagreb, Faculty of Electrical Engineering and Computing, Croatia;
Ministry of Education, Youth and Sports, MEYS  LM2023047, EU/MEYS CZ.02.1.01/0.0/0.0/16\_013/0001403, CZ.02.1.01/0.0/0.0/18\_046/0016007, CZ.02.1.01/0.0/0.0/16\_019/0000754, CZ.02.01.01/00/22\_008/0004632 and CZ.02.01.01/00/23\_015/0008197 Czech Republic;
CNRS-IN2P3, the French Programme d’investissements d’avenir and the Enigmass Labex, 
This work has been done thanks to the facilities offered by the Univ. Savoie Mont Blanc - CNRS/IN2P3 MUST computing center, France;
Max Planck Society, German Bundesministerium f{\"u}r Bildung und Forschung (Verbundforschung / ErUM), Deutsche Forschungsgemeinschaft (SFBs 876 and 1491), Germany;
Istituto Nazionale di Astrofisica (INAF), Istituto Nazionale di Fisica Nucleare (INFN), Italian Ministry for University and Research (MUR), and the financial support from the European Union -- Next Generation EU under the project IR0000012 - CTA+ (CUP C53C22000430006), announcement N.3264 on 28/12/2021: ``Rafforzamento e creazione di IR nell’ambito del Piano Nazionale di Ripresa e Resilienza (PNRR)'';
ICRR, University of Tokyo, JSPS, MEXT, Japan;
JST SPRING - JPMJSP2108;
Narodowe Centrum Nauki, grant number 2023/50/A/ST9/00254, Poland;
The Spanish groups acknowledge the Spanish Ministry of Science and Innovation and the Spanish Research State Agency (AEI) through the government budget lines
PGE2022/28.06.000X.711.04,
28.06.000X.411.01 and 28.06.000X.711.04 of PGE 2023, 2024 and 2025,
and grants PID2019-104114RB-C31,  PID2019-107847RB-C44, PID2019-104114RB-C32, PID2019-105510GB-C31, PID2019-104114RB-C33, PID2019-107847RB-C43, PID2019-107847RB-C42, PID2019-107988GB-C22, PID2021-124581OB-I00, PID2021-125331NB-I00, PID2022-136828NB-C41, PID2022-137810NB-C22, PID2022-138172NB-C41, PID2022-138172NB-C42, PID2022-138172NB-C43, PID2022-139117NB-C41, PID2022-139117NB-C42, PID2022-139117NB-C43, PID2022-139117NB-C44, PID2022-136828NB-C42, PDC2023-145839-I00 funded by the Spanish MCIN/AEI/10.13039/501100011033 and “and by ERDF/EU and NextGenerationEU PRTR; the "Centro de Excelencia Severo Ochoa" program through grants no. CEX2019-000920-S, CEX2020-001007-S, CEX2021-001131-S; the "Unidad de Excelencia Mar\'ia de Maeztu" program through grants no. CEX2019-000918-M, CEX2020-001058-M; the "Ram\'on y Cajal" program through grants RYC2021-032991-I  funded by MICIN/AEI/10.13039/501100011033 and the European Union “NextGenerationEU”/PRTR and RYC2020-028639-I; the "Juan de la Cierva-Incorporaci\'on" program through grant no. IJC2019-040315-I and "Juan de la Cierva-formaci\'on"' through grant JDC2022-049705-I. They also acknowledge the "Atracci\'on de Talento" program of Comunidad de Madrid through grant no. 2019-T2/TIC-12900; the project "Tecnolog\'ias avanzadas para la exploraci\'on del universo y sus componentes" (PR47/21 TAU), funded by Comunidad de Madrid, by the Recovery, Transformation and Resilience Plan from the Spanish State, and by NextGenerationEU from the European Union through the Recovery and Resilience Facility; “MAD4SPACE: Desarrollo de tecnolog\'ias habilitadoras para estudios del espacio en la Comunidad de Madrid" (TEC-2024/TEC-182) project funded by Comunidad de Madrid; the La Caixa Banking Foundation, grant no. LCF/BQ/PI21/11830030; Junta de Andaluc\'ia under Plan Complementario de I+D+I (Ref. AST22\_0001) and Plan Andaluz de Investigaci\'on, Desarrollo e Innovaci\'on as research group FQM-322; Project ref. AST22\_00001\_9 with funding from NextGenerationEU funds; the “Ministerio de Ciencia, Innovaci\'on y Universidades”  and its “Plan de Recuperaci\'on, Transformaci\'on y Resiliencia”; “Consejer\'ia de Universidad, Investigaci\'on e Innovaci\'on” of the regional government of Andaluc\'ia and “Consejo Superior de Investigaciones Cient\'ificas”, Grant CNS2023-144504 funded by MICIU/AEI/10.13039/501100011033 and by the European Union NextGenerationEU/PRTR,  the European Union's Recovery and Resilience Facility-Next Generation, in the framework of the General Invitation of the Spanish Government’s public business entity Red.es to participate in talent attraction and retention programmes within Investment 4 of Component 19 of the Recovery, Transformation and Resilience Plan; Junta de Andaluc\'{\i}a under Plan Complementario de I+D+I (Ref. AST22\_00001), Plan Andaluz de Investigaci\'on, Desarrollo e Innovación (Ref. FQM-322). ``Programa Operativo de Crecimiento Inteligente" FEDER 2014-2020 (Ref.~ESFRI-2017-IAC-12), Ministerio de Ciencia e Innovaci\'on, 15\% co-financed by Consejer\'ia de Econom\'ia, Industria, Comercio y Conocimiento del Gobierno de Canarias; the "CERCA" program and the grants 2021SGR00426 and 2021SGR00679, all funded by the Generalitat de Catalunya; and the European Union's NextGenerationEU (PRTR-C17.I1). This research used the computing and storage resources provided by the Port d’Informaci\'o Cient\'ifica (PIC) data center.
State Secretariat for Education, Research and Innovation (SERI) and Swiss National Science Foundation (SNSF), Switzerland;
The research leading to these results has received funding from the European Union's Seventh Framework Programme (FP7/2007-2013) under grant agreements No~262053 and No~317446;
This project is receiving funding from the European Union's Horizon 2020 research and innovation programs under agreement No~676134;
ESCAPE - The European Science Cluster of Astronomy \& Particle Physics ESFRI Research Infrastructures has received funding from the European Union’s Horizon 2020 research and innovation programme under Grant Agreement 

\end{document}